\documentclass[twocolumn,showpacs,preprintnumbers,amsmath,amssymb]{revtex4}

\usepackage{graphicx}
\usepackage{dcolumn}
\usepackage{bm}

\begin{document}


\title{Temporal behavior of synchronization between chaotic fiber lasers}

\author{Moti Fridman, Micha Nixon, Vardit Eckhouse, Nir Davidson, and Asher A. Friesem}

\affiliation{Dept. of Physics of Complex system, Weizmann
institute of Science, Rehovot 76100, Israel}

\date{\today}

\begin{abstract}
A new configuration for synchronizing two chaotic fiber lasers,
which includes both coupling and losses, is presented.
Experimental and calculated results reveal that the
synchronization time can be significantly shorter than with
configuration that have only coupling.
\end{abstract}

\maketitle

In general, chaotic lasers have received considerable attention in
the past few years, because they are potentially useful for
applications such as encryption and secure
communication~\cite{DiodeLasersReview, ChaoticLasersNature,
RoyComS}. More recent, attention has been shifted to chaotic fiber
lasers that have a rich and very broad
bandwidth~\cite{ChaoticLasersNature, RoyComS, TangCom,
KusumotoCom, RoyCom, AbarbanelChaoticFiberLaser, FiberLasersLuo,
FiberLasersKim}. In these applications synchronization plays a
dominant role where the most crucial factor is the time that is
necessary for synchronizing two or more chaotic
lasers~\cite{PorteDiodeLaser, KantorPRL, FiberLasersLuo,
FiberLasersKim}.

In this letter, we investigate the temporal dynamics of
synchronization between two chaotic fiber lasers. We consider two
different configurations. One is a modified version of the
configuration which is normally used for synchronizing two lasers,
in which the coupling between the lasers is performed outside
their cavity - namely outer-cavity configuration. The other
configuration is one in which the coupling between the lasers is
performed inside their combined cavity - namely intra-cavity
configuration. We develop relatively simple models for both
configurations that predict that synchronization will occur much
faster with the intra-cavity configuration, in agreement with the
experimentally obtained results.

\begin{figure}[h]
\includegraphics[width=8.3cm]{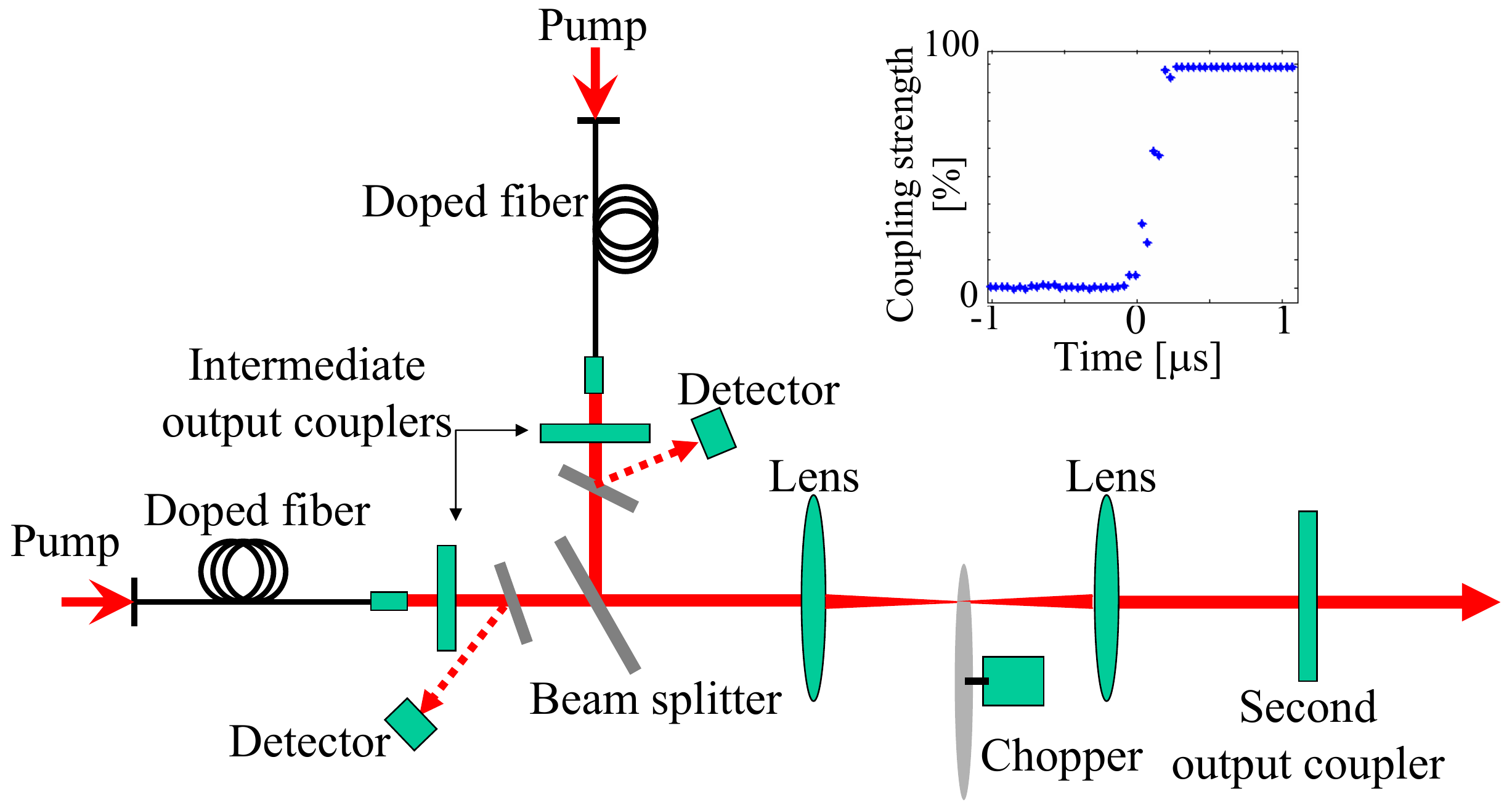}
\caption{\label {System}(color online) Basic outer-cavity
configuration for synchronizing between two chaotic fiber lasers
and measuring the temporal behavior.}
\end{figure}

The basic outer-cavity configuration is presented schematically in
Fig.~\ref{System}. It includes two fiber lasers with an
intermediate output coupler, a 50/50 beam splitter and an
additional combined output coupler for coupling and synchronizing
the two lasers. Also included is a two lens 4f optical arrangement
in which a chopper which spins at $1000Rpm$ is placed at the focal
plane between the lasers for rapidly switching on and off the
coupling between the lasers. With such an arrangement it was
possible to experimentally obtain either no coupling or strong
coupling with coupling strength rise time of $120 ns$, as shown in
the inset. Each fiber laser consists of Erbium doped fiber of
about ten meters in length, where one end is attached to a high
reflection fiber Bragg grating (FBG) with central wavelength of
$1550 nm$ and bandwidth of about $1nm$ that serves as a back
reflector mirror and the other end is spliced to a collimating
graded index (GRIN) lens with anti-reflection layer to suppress
any reflections back into the fiber cores, and an intermediate
output coupler of 20\% reflectivity. Each fiber laser is pumped
with a rapidly oscillating diode laser which imposes chaotic
behavior, to obtain two chaotic fiber lasers. The beam splitter
combines the beams from the two chaotic lasers into one beam and
the additional output coupler reflects back part of the light of
the combined beam back into the fiber lasers, thereby obtaining
coupling between them~\cite{VarditPRL, Moti2006}.

The intra-cavity configuration is similar to that shown in
Fig.~\ref{System}, but without the intermediate output couplers.
Accordingly, the beam splitter is now inside the cavities of the
lasers and the combined output coupler is common for both fiber
lasers. In this configuration, the beam splitter introduces high
losses to each of the lasers when they are not synchronized, i.e.
the losses depend on the synchronization. In general, the lasers
prefer to operate at minimum losses. The minimization of losses
occurs very rapidly, so the two lasers will synchronize even at a
much faster rate than at the outer-cavity configuration, were
there are no such losses at all.

\begin{figure}[h]
\centerline{\includegraphics[width=8.3cm]{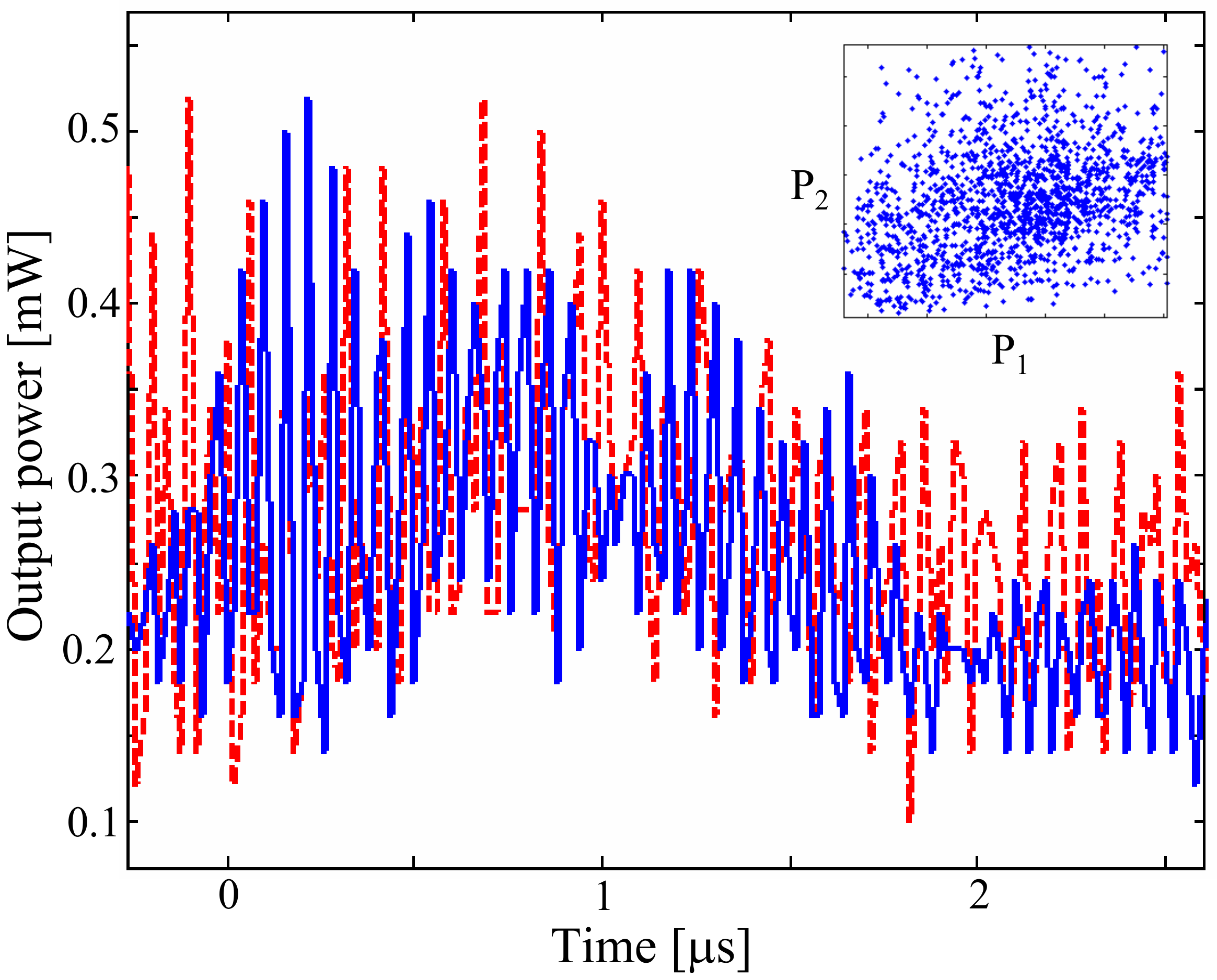}}
\caption{Experimental results of the output power of the two
lasers as a function of time when the lasers are not synchronized.
Inset shows the output power of one fiber laser as a function of
the output power of the other.} \label {NotSynch}
\end{figure}

\begin{figure}[h]
\centerline{\includegraphics[width=8.3cm]{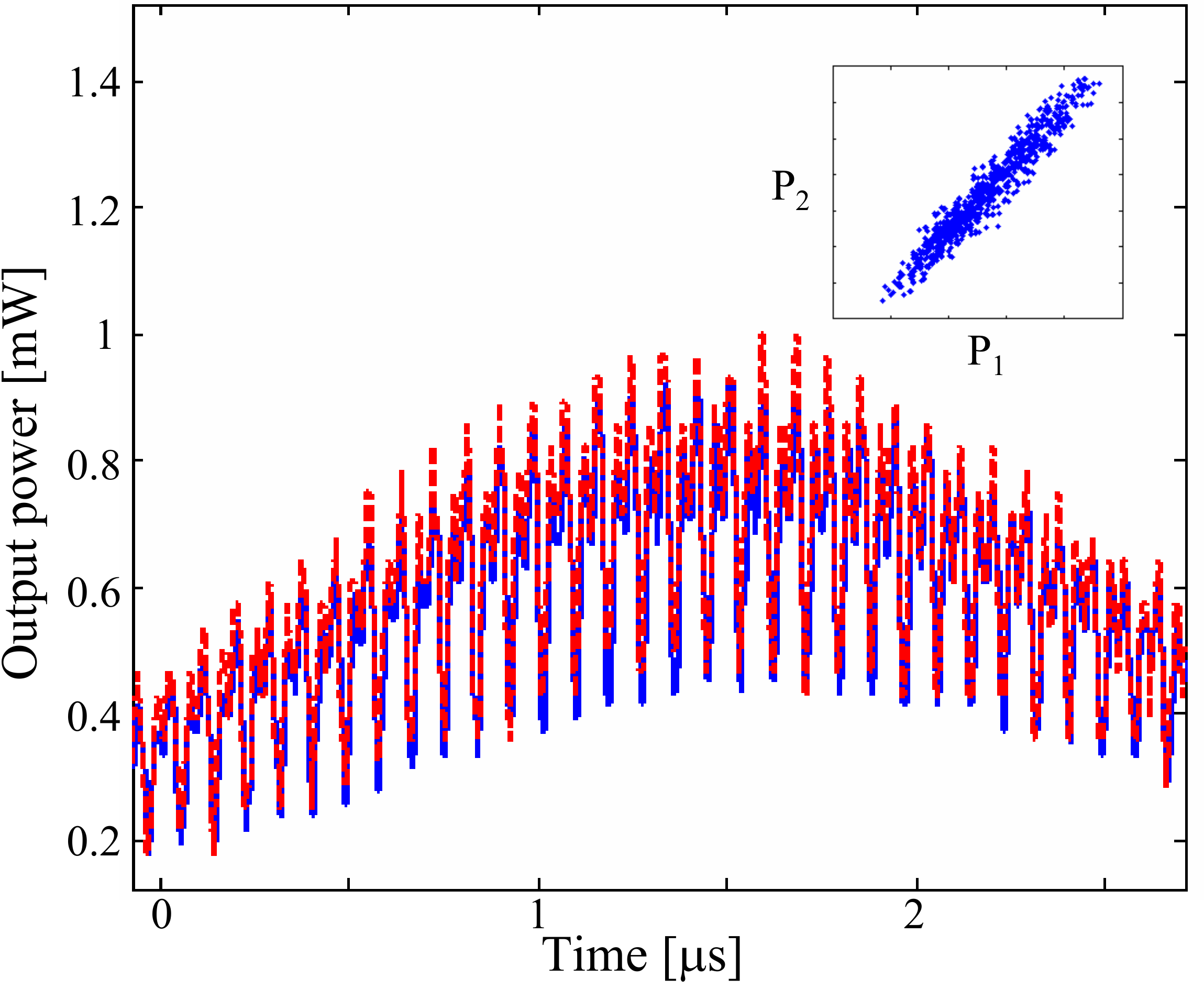}}
\caption{(color online) Experimental results of the output power
of the two lasers as a function of time when the lasers are
synchronized. Inset shows the output power of one fiber laser as a
function of the output power of the other.} \label {Synch}
\end{figure}

We began our experiments by simultaneous detecting the output
power as a function of time for each laser. This was done by
reflecting a small portion of the light that is propagating in
each lasers towards fast photo-detectors. Representative results
are shown in Figs.~\ref{NotSynch} and ~\ref{Synch}. Figure
\ref{NotSynch} shows the results when the lasers are not coupled.
As expected, the two signals are not synchronized. This was
confirmed by plotting the output power of one fiber laser as a
function of the output power of the other at the inset in
Fig.~\ref{NotSynch}. The wide spread of points indicates that
there is almost no synchronization between the lasers. Indeed, we
calculated the corresponding correlation coefficient and found it
to be as low as 0.15.

Figure \ref{Synch} shows the results when the lasers are strongly
coupled. As evident, the lasers are now synchronized. This is
confirmed by plotting the output power of one laser as a function
of the output power of the other, at the inset in
Fig.~\ref{Synch}, which yield a narrow linear distribution. The
corresponding calculated correlation coefficient for this
distribution was over 0.98. It should be noted that the intensity
fluctuation of the lasers in Figs.~\ref{NotSynch} and ~\ref{Synch}
are mainly due to beating between different longitudinal modes.
Thus, the synchronization between the fiber lasers in our
configurations is not only due to synchronization of intensity
fluctuation but also due to phase locking of the lasers
longitudinal modes.

We now determined the temporal behavior of synchronization between
the two lasers around the transition from no coupling to full
coupling. This was done by measuring simultaneously the output
powers as a function of time for the two lasers around the
transition region and calculating their correlation coefficient
within a moving window of $100 ns$ width~\cite{KantorPRL}. This
correlation coefficient is essentially the instantaneous
synchronization between the lasers.

\begin{figure}[h]
\centerline{\includegraphics[width=8.3cm]{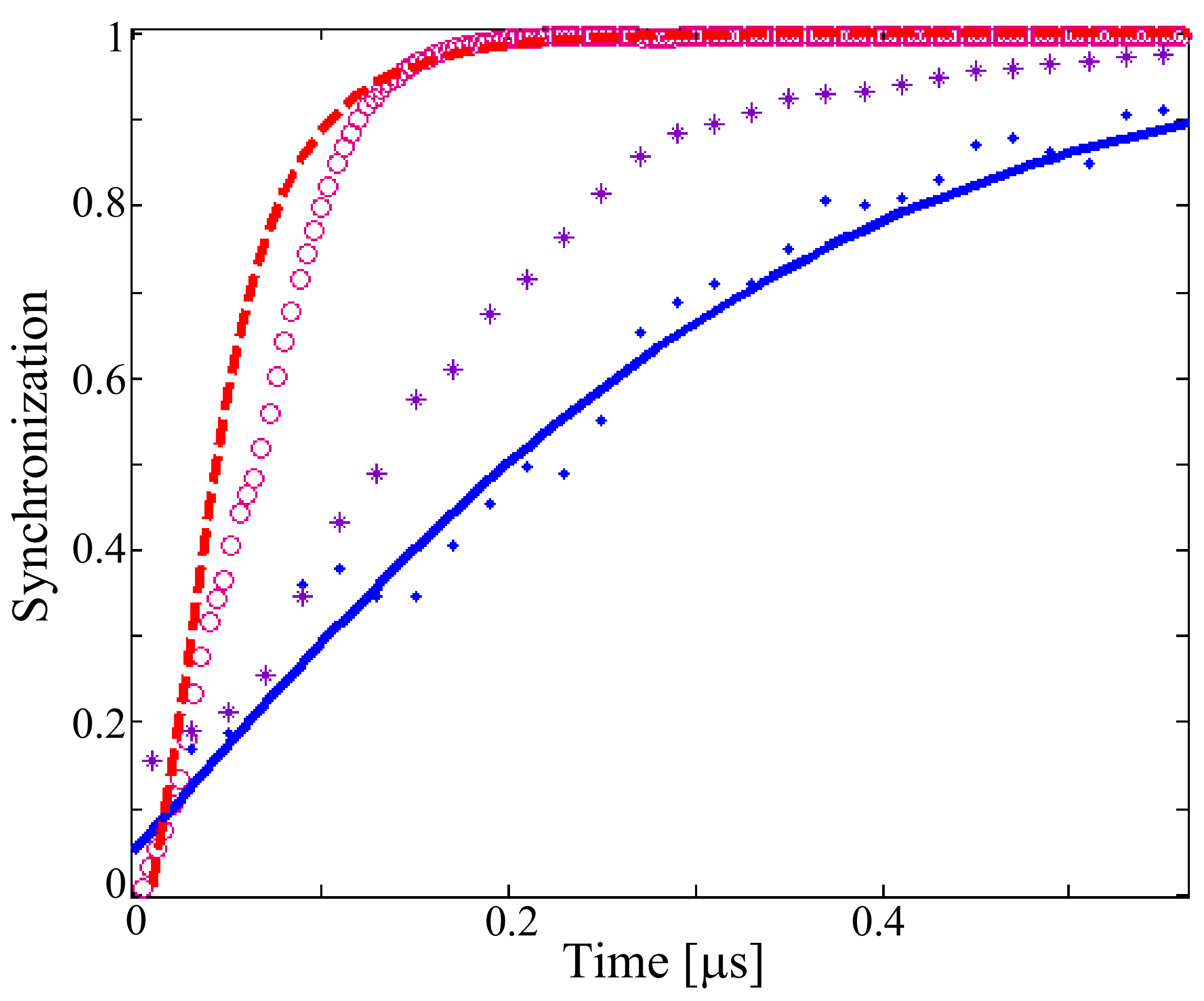}}
\caption{(color online) Experimental and calculated results of the
synchronization between two chaotic fiber lasers as a function of
time after the onset of coupling. Dots denote experimental results
for the outer-cavity configuration. Circles denote experimental
results for the intra-cavity configuration. Stars denote
experimental results for configuration between the outer-cavity
and the intra-cavity configurations. Solid curve denotes
calculated results for the outer-cavity configuration. Dashed
curve denotes calculated results for the intra-cavity
configuration.} \label {AllResults}
\end{figure}

The results of synchronization as a function of time starting at
the onset of coupling for both the outer-cavity and intra-cavity
configuration, are shown in Fig.~\ref{AllResults}. The dots denote
the experimentally obtained results for the outer-cavity
configuration, indicating that it takes $350 ns$ to reach 80\%
synchronization. The circles denote the experimentally obtained
results for the intra-cavity configuration, indicating that it
takes $100 ns$ to reach 80\% synchronization, probably limited by
the finite rise time of the coupling as dictated by the chopper
optical arrangement. These results clearly show that it is at
least three times faster to reach reasonable synchronization with
the intra-cavity configuration. To elucidate these results we also
present experimental results for configuration which has
intermediate output coupler with much lower reflection than in the
outer-cavity configuration but still non-zero as it was in the
intra-cavity configuration. These results are presented in
Fig.~\ref{AllResults} as the starts. In this intermediate
configuration, the time in takes for the synchronization to reach
80\% is $200 ns$, which is indeed in-between the two extreme
cases.

In order to support our results, we developed a model to help
elucidate the temporal behavior of synchronization between two
fiber lasers. To analyze the complex behavior of fiber lasers one
should consider a detailed model~\cite{Pisarchik}, introducing the
parameters of our system this model is reduced to the well known
equations of coupled single longitudinal mode
lasers~\cite{Fabiny}, as
\begin{equation}
\frac{dE_{1,2}}{dt}=\frac{1}{\tau_{c}}\left[\left(G_{1,2}-\alpha\right)E_{1,2}+\kappa
E_{2,1}\right]+i \omega_{1,2} E_{1,2}, \label{Fields}
\end{equation}

\begin{equation}
\frac{dG_{1,2}}{dt}=\frac{1}{\tau_{f}}\left(P-G_{1,2}-G_{1,2}|E_{1,2}|^{2}\right),
\label{Gain}
\end{equation}
where E is the electric field in each of the lasers, G the gain,
$\tau_{c}$ the life time of the cavity, $\tau_{f}$ the life time
of the excited state in the gain, $\alpha$ the losses in the
cavities, P the pumping rate, $\kappa$ the coupling strength
between the lasers and $\omega$ the frequency in each laser. Then
we substitute $E_{1,2}=A_{1,2} \exp (\imath \varphi _{1,2})$ with
$A_{1,2}$ the amplitude of each laser, and define $\beta$ as the
ratio between these amplitudes. as
\begin{equation}
\beta=\frac{A_{1}}{A_{2}}. \label{betadef}
\end{equation}
where $d \beta / dt$ is a measure for the synchronization, i.e. $d
\beta / dt=0$ indicates that the two lasers are synchronized.

In order to analytically evaluate the temporal behavior of
synchronization between the lasers in the outer-cavity
configuration, we use Eqs.~(\ref{Fields}) and ~(\ref{betadef}) to
derive an equation of motion for $\beta$, as
\begin{equation}
\frac{d\beta}{dt}=\frac{\kappa}{\tau_{c}}\left(1-\beta^{2}\right)\cos\left(\varphi_{1}-\varphi_{2}\right).
\label{betamotion}
\end{equation}
Thus, when the two lasers are synchronized $\beta=1$. Since phase
locking occurs very rapidly~\cite{FastPhaseLocking}, so
$\varphi_{1}-\varphi_{2}<1$, we can neglect the $\cos$ term. Thus,
the solution of Eq.~(\ref{betamotion}) is
\begin{equation}
\beta(t)=\frac{\exp\left(\frac{2 \kappa
t}{\tau_{c}}\right)+C}{\exp\left(\frac{2 \kappa
t}{\tau_{c}}\right)-C}, \label{betaSol}
\end{equation}
where C is a constant of integration that is set by the initial
conditions. Using Eq.~(\ref{betaSol}) we calculated $\beta$ as a
function of time. The results are presented as the solid curve
(blue) in Fig.~\ref{AllResults}. As evident, there is a good
agreement between the experimental results and those predicted by
the model, indicating that the assumption of
$\varphi_{1}-\varphi_{2}<1$ was justified.

For the intra-cavity configuration, the cavity losses are reduced
when there is destructive interference between the two beams in
the loss channel. The efficiency of the destructive interference
is a function of the relative phase and the relative intensity of
the two lasers. Assuming that the phase locking accurse much
faster than the synchronization, the losses are a function of only
$\beta$. Accordingly, in order to analytically evaluated the
temporal behavior of the synchronization between the lasers in the
intra-cavity configuration we replaced the constant losses
$\alpha$ in Eq.~(\ref{Fields}) with a losses which are a function
of $\beta$, as
\begin{equation}
\alpha(\beta)=\frac{\left(\beta-1\right)^{2}}{2}. \label{lossDef}
\end{equation}
Now we use Eqs.~(\ref{Fields}), ~(\ref{betadef}) and
~(\ref{lossDef}) to derive the equation of motion of $\beta$, as
\begin{equation}
\frac{d\beta}{dt}=\frac{1}{\tau_{c}}\left(\frac{\left(\beta-1\right)^{2}
\beta
}{2}+\left(1-\beta^{2}\right)\kappa\cos\left(\varphi_{1}-\varphi_{2}\right)\right).
\label{betamotionloss}
\end{equation}
Again assuming $\varphi_{1}-\varphi_{2}<1$ and also assuming a
small coupling of $\kappa<0.1$, then an approximate analytic
solution of Eq.~(\ref{betamotionloss}) is
\begin{equation}
\beta(t)=\frac{\exp\left(\frac{6 \kappa
t}{\tau_{c}}\right)+C}{\exp\left(\frac{6 \kappa
t}{\tau_{c}}\right)-C}, \label{betaSolLoss}
\end{equation}
where C is a constant of integration that is set by initial
conditions. Using Eq.(\ref{betaSolLoss}), where exponentials are
three times larger than these of Eq.(\ref{betaSol}), we calculated
the evolution of $\beta$ as a function of time. The results are
presented as the dashed curve (red) shown in
Fig.~\ref{AllResults}. As evident, the experimentally obtained
rise-time is somewhat longer than the predicted, probably because
it is limited by the finite rise-time of the coupling which is
about $120ns$.

To conclude, we investigated the temporal behavior of
synchronization between two chaotic fiber lasers when the coupling
between them is suddenly switched on. Two configurations were
considered, one where the synchronization is achieved with just
coupling and the other by coupling as well as additional losses.
The results reveal that with the additional losses the rise-time
was about three times shorter than with just coupling. We
performed our experiment for a wide range of pump modulation
parameters, some of them chaotic and some not, and obtained very
similar results for both regimes. Such reduction of rise time
should be useful in number of applications such as higher possible
bit rate in secure communication.


\end{document}